\documentclass[fleqn,usenatbib]{mnras}

\usepackage{newtxtext,newtxmath}

\usepackage[T1]{fontenc}
\usepackage{ae,aecompl}

\usepackage{graphicx}	
\usepackage{amsmath}	
\usepackage{amssymb}	
\usepackage{siunitx}
\usepackage{subcaption}

\newcommand{\abs}[1]{\ensuremath{\left\lvert {#1}\right\rvert}}
\newcommand{\chsqr}{\ensuremath{\chi^2}}

\newcommand{\lcdm}{\ensuremath{\Lambda\text{CDM}}}

\newcommand{\dlum}{\ensuremath{d_\text{L}}}

\newcommand{\deriv}[2]{\ensuremath{\frac{\diff {#1}}{\diff {#2}}}}

\newcommand{\fseight}{\ensuremath{f\sigma_8}}
\newcommand{\hMpc}[1]{\ensuremath{#1 \, h^{-1} \mathrm{Mpc}}}

\newcommand{\Omm}{\ensuremath{\Omega_\text{m}}}
\newcommand{\Omo}{\ensuremath{\Omega_\text{m,0}}}

\newcommand{\Omde}{\ensuremath{\Omega_\text{de}}}

\newcommand{\diff}{\ensuremath{\mathrm{d}}}

\newcommand{\vect}[1]{\ensuremath{\mathbfit{#1}}}
\newcommand{\tens}[1]{\ensuremath{\mathbfss{#1}}}

\newcommand{\seighto}{\ensuremath{\sigma_{8,0}}}
\newcommand{\hrec}{\ensuremath{\hat{h}}}
\newcommand{\Geff}{\ensuremath{G_\text{eff}}}
\newcommand{\fseightrec}{\ensuremath{\widehat{\fseight}}}
\newcommand{\zmax}{\ensuremath{z_\text{max}}}

\title[Defying the laws of Gravity I]{Defying the laws of Gravity I: Model-independent reconstruction of the Universe expansion from growth data}

\author[B. L'Huillier et al.]{
Benjamin~L'Huillier,$^{1,2}$
Arman~Shafieloo,$^{2,3}$
David~Polarski,$^4$
Alexei~A.~Starobinsky$^{5,6}$
\\
$^{1}$Department of Astronomy, Yonsei University, Yonsei-ro 50, Seodaemun-gu, Seoul 03722, Korea\\
$^{2}$Korea Astronomy and Space Science Institute, Yuseong-gu, Daedeok-daero 776, Daejeon 34055, Korea\\
$^{3}$University of  Science and Technology,  Yuseong-gu 217 Gajeong-ro, Daejeon 34113, Korea\\
$^{4}$Laboratoire Charles Coulomb, Universit\'e de Montpellier \& CNRS UMR 5221,F-34095 Montpellier, France\\
$^{5}$L. D. Landau Institute for Theoretical Physics RAS, Moscow 119334, Russia\\
$^{6}$National Research  University Higher  School of  Economics, Moscow 101000, Russia
}

\date{Accepted 2020 February 5. Received 2020 January 31; in original form 2019 July 8}

\pubyear{2019}

\begin{document}
\label{firstpage}
\pagerange{\pageref{firstpage}--\pageref{lastpage}}
\maketitle

\begin{abstract}
Using redshift space distortion data, we perform model-independent reconstructions of the growth history of matter inhomogeneity 
in the expanding Universe using two methods: crossing statistics and Gaussian processes. 
We then reconstruct the corresponding history of the Universe background expansion and fit it to type Ia supernovae data, putting  constraints on $(\Omo,\seighto)$. The results obtained are consistent with the concordance flat-\lcdm\ model and General Relativity as the gravity theory given the current quality of the inhomogeneity growth data. 
\end{abstract}




\section{Introduction}
The discovery at the end of last century of the late-time accelerated expansion rate of the Universe raised the question of its physical cause. There are two great avenues towards solving this problem: either that it is due to an unknown new physical component dubbed (physical) dark energy, or its origin lies in a modification of the laws of gravity \citep[e.g.,][]{Sahni:1999gb,2012IJMPD..2130002Y,2012PhR...513....1C}. However, they represent particular cases of a more general situation like in scalar-tensor gravity, when both a new physical field is introduced for dark energy description and gravity is modified, too (see e.g. \citet{Boisseau:2000pr,Copeland:2006wr,Sahni:2006pa}).

In the concordance model the role of dark energy is played by a cosmological constant $\Lambda$ while gravity is described by Einstein's theory of General Relativity (GR). While GR has been remarkably successful to explain observations in the Solar system (see e.g. \citet{2006JCAP...09..016G}), its successful extrapolation to much larger cosmic scales remains unclear. Modified gravity models with modifications of gravity on cosmic scales are not excluded and could well be the solution to the recent acceleration of the Universe.  The nature of dark energy and therefore also the correct model of gravity are burning issues of cosmology and theoretical physics in general. 

The large-scale structures of the Universe are an ideal laboratory to test gravity, and to distinguish between physical dark energy and modified gravity (which may be also called geometrical dark energy as in \citet{Sahni:2006pa}). In particular, redshift-space distortion (RSD) due to galaxy peculiar velocities can be used to estimate the growth factor $f$, which is a key to understanding gravity. 
However, as a first step, we restrict ourselves to GR. 
For a flat-FLRW Universe  with dark energy as a perfect fluid with equation of state $w(z)$, the expansion history $h(z) = H(z)/H_0$ is described by
\begin{align}
    \label{eq:expans}
    h^2(z) &  = \Omo(1+z)^3 + (1-\Omo)\exp\left(3\int_0^z \frac{1+w(u)}{1+u}\diff u\right).
\end{align}
In GR, the evolution of the matter overdensity $\delta(\vect x,z) = (\rho(\vect x,z)-\bar\rho(z))/\bar\rho(z)$ are governed in the Newtonian approximation by
\begin{align}
\label{eq:growth}
\ddot\delta + 2H\dot\delta & = \frac 3 2 H^2 \Omm \delta,
\end{align}
where dot stands for a derivative with respect to cosmic time $t$, and $\Omm(z) = \Omo (1+z)^3/h^2(z)$ is the matter density normalized by the critical density.
From eqs.~\eqref{eq:expans} and ~\eqref{eq:growth}, it is clear that changing the expansion will also affect the growth evolution. 
In fact, \citet{1998JETPL..68..757S} showed that the Universe expansion history $H(z)$ can be also reconstructed from  $\delta(z)$ unambiguously in this case (the situation becomes more complicated in scalar-tensor gravity \citep{Boisseau:2000pr}).

In this paper, we aim first to reconstruct  the growth history from data, and then to use it in order to deduce the expansion history (assuming GR) and to compare it with the supernovae data. 
We should note here that there are two different reconstructions involved here: a statistical reconstruction of the growth factor $f(z)$ from observational data on one hand, and on the other hand, the theoretical reconstruction of the background expansion $H(z)$ from $f(z)$.
\citet{2014JCAP...02..021L} applied the \citet{1998JETPL..68..757S} formalism by integrating from $z$ to a maximum redshift $z_\mathrm{max}$ and henceforth obtain $\Omo$, and studied the effect of $\zmax$ on the error budget, showing it dominates over the data uncertainties. 
However, this approach differs from ours in that we integrate from $0$ to $z$ and use $\Omo$ as a free parameter. 
A more similar approach to ours was recently applied in \citet{2018arXiv180800377Y}.
The theoretical reconstruction is described in \S~\ref{sec:method}, and the statistical reconstructions together with the results are presented in \S~\ref{sec:real}, and our conclusions are drawn in \S~\ref{sec:ccl}. 
We validate the method on mock data in \S~\ref{sec:mock}.

\section{Method}
\label{sec:method}

\subsection{Theoretical Framework}

From eq.~\eqref{eq:growth}, using that for any function $x(t)$,
\begin{align}
\dot x &=  H \deriv x {\ln a}=-H\deriv x {\ln (1+z)},
\intertext{one obtains}
\deriv{^2\delta}{\ln a^2} & + \left(2+\frac{1}{h(z)}\deriv{h}{\ln a}\right) \deriv\delta{\ln a} = \frac 3 2 \Omm(z) \delta.
\intertext{It is convenient to introduce the growth factor}
f & = \deriv {\ln \delta}{\ln a} =  \Omm^\gamma(a), \label{eq:gamma}
\end{align}
where the last equality defines the growth index $\gamma$.
In general, $f=f(\vect k,z)$ and therefore $\gamma = \gamma(\vect k,z)$ \citep[e.g.,][]{2009JCAP...02..034G}. 
However, in GR and for dust-like matter, $\gamma$ is $\vect k$-independent and has weak dependence on $z$, so that $\gamma(z)\simeq 0.55$, with a slight dependence on the equation of state of dark energy $w$ and the matter density parameter \Omo. Note, however, that 
$\gamma$ may not be exactly $z$-independent for quintessence (a scalar field with a potential minimally coupled to gravity) models of dark energy as was shown in  \citet{2016JCAP...12..037P}.

Observational data on RSD provide us with the product \fseight, where 
\begin{align}
\fseight & = \deriv{\sigma_8}{\ln a}\\
\intertext{where}
\sigma^2_R(z) & = \frac 1 {2\pi^2}\int_0^\infty P(k,z) W^2_R(k) k^2\diff k \propto\delta^2(z)
\end{align}
is the rms of the density fluctuations smoothed over a radius $R$, usually taken to be \hMpc{8}. 

In \citet{2018PhRvD..98h3526S} \citep[see also][]{2018MNRAS.476.3263L}, we used Pantheon and a compilation of growth data to put model-independent constraints on $(\Omo,\seighto,\gamma)$, where subscript 0 stands for the current value.
However, in that paper, we treated $\gamma$ as a constant, effectively performing a consistency test of GR.  
In fact, one can solve the problem without assuming $\gamma = $ constant.
Assuming $\delta$ and $\delta'$ are known, the expansion history $h$ can be uniquely determined via \citep{1998JETPL..68..757S} 
\begin{align}\label{eq:delta2h2}
 h^2(z) & = \left(\frac{1+z}{\delta'(z)}\right)^2 \left(
 \delta'^2_0-3\Omo\int_0^z\delta(u)\abs{\delta'(u)}\frac{\diff u}{1+u}\right),
\end{align}
where $'$ denotes a derivative with respect to  $z$ (not $\ln a$).

Thus, one can obtain from RSD measurements 
\begin{align}
\frac{\delta'(z)}{\delta_0} &= -(1+z)\frac {\fseight(z)}{\seighto}, \mbox{and}\\
\delta(z) & = \delta_0\left(1-\frac 1{\seighto}\int_0^z \fseight(u)\frac{\diff u}{1+u}\right). 
\end{align}

Therefore, for a given reconstruction $\fseightrec(z)$ together with a given $(\Omo,\seighto)$, the expansion history $\hrec(z)$ is uniquely determined by
\begin{multline}
    \hrec^2(z|\fseightrec,\Omo,\seighto)  = \frac{(1+z)^4}{\fseightrec^2(z)}
\left({\fseightrec^2_0}\phantom{\int}\right.\\
\left.
-3{\Omo}
\int_0^z\left(\seighto-
\int_0^u \fseightrec(v)\frac{\diff v}{1+v}\right)
{\fseightrec(u)}
\frac{\diff u}{(1+u)^2}
\right).
\label{eq:h2}
\end{multline}

Therefore, reconstructing $\fseightrec(z)$, and exploring the $(\Omo,\seighto)$ parameter space, we can reconstruct $h(z)$ and the luminosity distance. 
In the following, we assume a flat-FLRW universe, which is consistent with current model-dependent \citep{2013PhLB..723....1F,2018arXiv180706209P} and model-independent\citep{2014JCAP...03..035R, 2015PhRvL.115j1301R,  2017JCAP...01..015L, 2018JCAP...03..041D, 2018PhRvD..98h3526S} constraints.
The luminosity distance is thus
\begin{align}
  \dlum(z) & = \frac c {H_0} (1+z) \int_0^z\frac{\diff x}{h(x)},\\
\intertext{and the corresponding distance modulus}
  \mu(z) & = 5 \log_{10}(\dlum / \SI{1}{Mpc}) + 25,
\end{align}
we can then fit the reconstructed $\mu$ to SNIa data, obtain a \chsqr, and hence put model-independent constraints on $(\Omo,\seighto)$.

\subsection{Fit to the data}

After obtaining $\fseight$ from GP, we then obtain the expansion history $h$ via eq.~\eqref{eq:delta2h2}, for a given choice of $(\Omo,\seighto)$. 
We note that, a priori, for a given $(\fseight,\Omo,\seighto)$, there is no guarantee for $h^2(z)$ to be larger than the matter term $\Omo(1+z)^3$. 
At low redshifts, local measurements yield $\Omo<1$, and thus,  in the absence of spatial curvature, the energy density of dark energy is positive. 
However, the energy density of dark energy could be negative in the matter-dominated era.
An effective negative $\rho_\text{de}$ 
was considered in e.g. \citet{2015JCAP...07..002B}.
Following these considerations, hereafter, we impose the positive dark energy condition that
\begin{align}
    \label{eq:pde}
    \Omde(z) & = 1 -\Omm(z) \geq 0 \qquad \forall z<\zmax
\end{align}
In addition, even when $\Omde<0$, we impose the conservative choice $\Omde>-0.05$.
If we assume tracking dark energy during the matter era, CMB constraints which fix the perturbations amplitude at $z\approx 1100$ give a slightly stronger lower bound $\Omde>-0.025$ if we restrict the relative additional growth to no more than $10\%$.   
Indeed, we get for constant $\left|\Omde\right|\ll 1$ during the matter dominated stage
\begin{align}
\delta(z) \propto (1+z)^{\frac14 \left[ 1 - \sqrt{ 25 - 24\Omde} \right] } \approx (1+z)^{-1+\frac35 \Omde}~.
\end{align}
This anomalous growth is reminiscent of the growth in the presence of massive neutrinos (where there is a weaker growth with expansion) with the essential difference that here the growth is boosted due to a negative $\Omde$.

In order to study the effect of $\zmax$, we choose three values for ${\zmax} = $ 0.7, 1, and 2.
We note that $w$ will then diverge when $\Omde$ crosses zero (except for the degenerate case where the DE pressure also vanishes at the same moment). 

Finally, calculate the \chsqr\ for all the data:
\begin{align}
    \chsqr_\text{tot} & = \chsqr_{\fseight}+\chsqr_\text{SNIa}.
\end{align}

\begin{figure*}
    \centering
    \includegraphics[width=\textwidth]{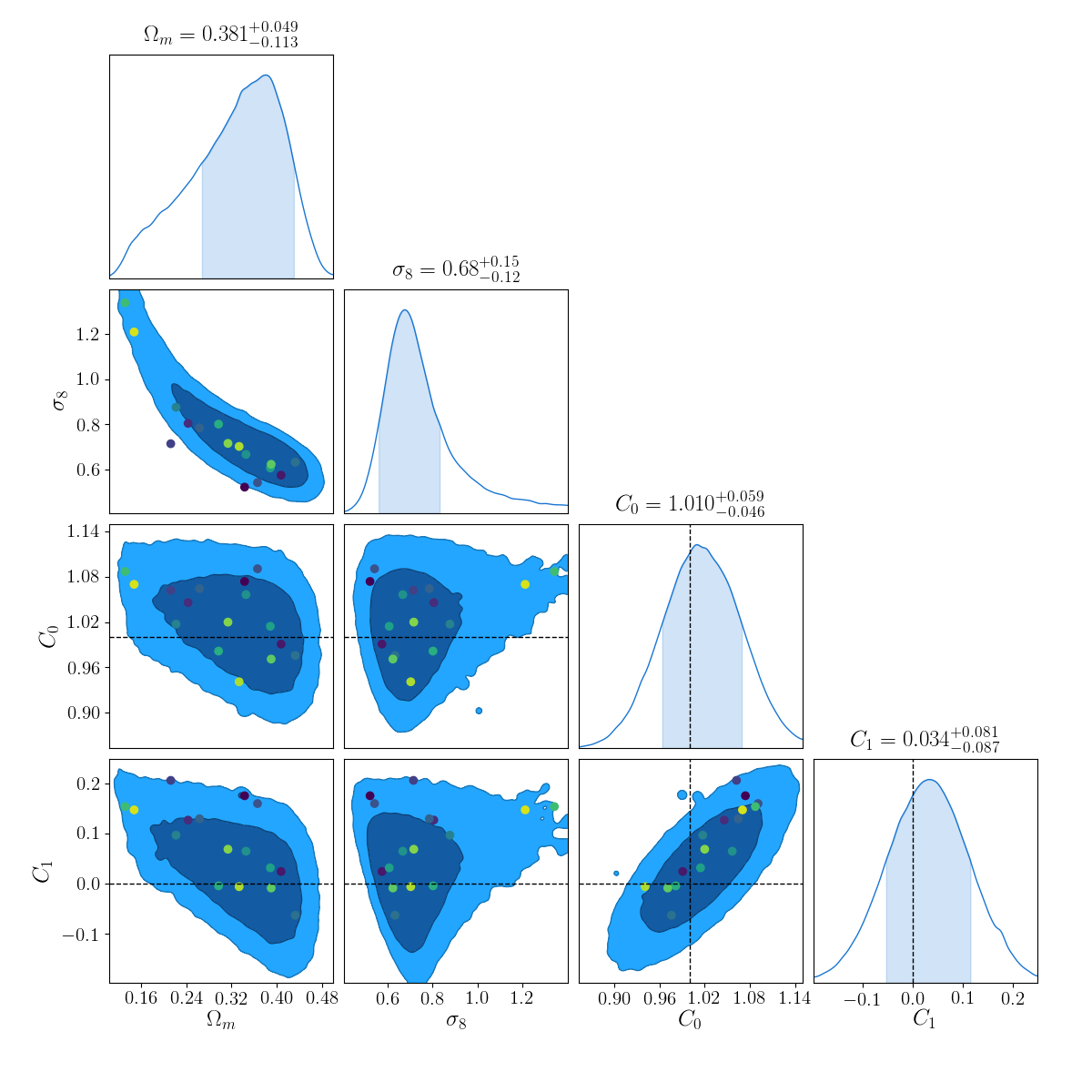}
    \caption{68\% and 95\% confidence area of the posterior of the cosmological parameters $(\Omo,\seighto)$ and the Chebyshev coefficients. The dashed lines show the $C_0=1$ and $C_1=0$, i.e., no deviation from the best-fit }
    \label{fig:mcmc_xing}
\end{figure*}

\section{Results}
\label{sec:real}

We used the same data sets as in \citet{2018PhRvD..98h3526S}: the Pantheon compilation \citep{2018ApJ...859..101S}, and the compilation of RSD data including: 
2dFGRS \citep{2009JCAP...10..004S},
WiggleZ \citep{2011MNRAS.415.2876B},
6dFGRS \citep{2012MNRAS.423.3430B},
VIPERS \citep{2013MNRAS.435..743D},
the SDSS Main galaxy sample \citep{2015MNRAS.449..848H}, 
2MTF \citep{2017MNRAS.471.3135H}, 
BOSS DR12 \citep{2017MNRAS.465.1757G},
FastSound \citep{2016PASJ...68...38O},  and
eBOSS DR14Q \citep{2019MNRAS.482.3497Z}.

\subsection{Crossing statistics}

\begin{figure*}
    \centering
    \includegraphics[width=\textwidth]{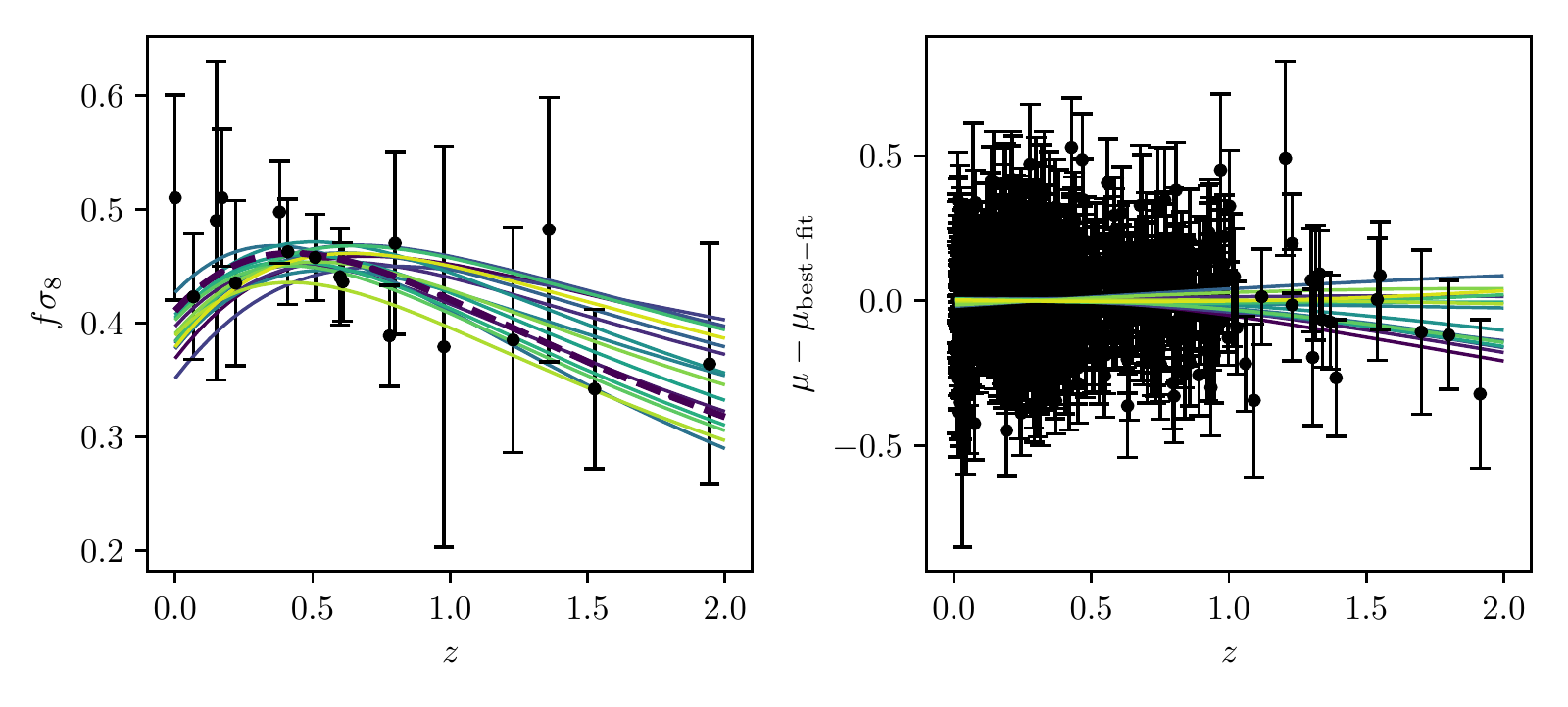}
    \caption{Data, best-fit (dashed lines), and 16 random choices of the $(\Omo,\seighto,C_0,C_1)$ MCMC sampling (solid lines) and their associated \fseightrec (left) and $\hat\mu$ (right).}
    \label{fig:crossing}
\end{figure*}

In this section, we study the effects of distorting the mean function on the final fit. 
In practice, this is equivalent to applying the Bayesian interpretation of the crossing statistics formalism to the reconstructed growth history \citep{2011JCAP...08..017S, 2012JCAP...05..024S, 2012JCAP...08..002S}.
In this formalism, the prediction from the theory to be tested, \lcdm+GR in the present case, is multiplied by  some hyperfunction $T_N(x|C_0,\dots,C_N)$: 
\begin{align}
\fseightrec(x) &  = (\fseight)^{\mathrm{\lcdm}}(x) \times T_N(x|C_0,\dots,C_N),
\intertext{where}
T_N(x|C_0,\dots,C_N) &= \sum_i^N C_i P_i(x), \\
x & = 2\left(\frac{z}{z_\mathrm{max}}-\frac 1 2\right),
\end{align}
$P_i(x)$ is the $i$th order Chebyshev polynomial of the first kind, and $C_i$ are free parameters.
The Chebyshev polynomials constitute an orthogonal basis for $x\in[-1,1]$, and as such, can represent any function. 
The zeroth order controls the absolute scaling, the first order the tilt, and higher order introduce a curvature and  inflexion points. 
In the Bayesian interpretation of the crossing statistics, we are interested in the confidence intervals around the hyperparameters $C_i$. 
If the $C_i$ are consistent with $C_0=1, C_i=0 (i\geq 1)$, the data have no preference for any departure from the mean function, meaning the model is consistent with the data. 
In case of significant deviation from $C_0=1,C_i=0 (i\geq 1)$, the data suggest a preferred deformation of the mean function.

Distorting the starting $\fseight$, we obtain $\hat\mu$ and fit both $\hat\mu$ and \fseightrec\ to the data.
As noted by \citet{2014PhRvD..89d3004H}, going towards too high orders, one might miss the effects of the lower orders. 
Therefore, we start by limiting to the first order, i.e., tilting the mean function.
We used the \texttt{emcee} Monte-Carlo Markov Chain (MCMC) package \citep{2013PASP..125..306F} to explore the parameter space $(\Omo, \seighto,C_0,C_1)$, and show the posteriors in Fig.~\ref{fig:mcmc_xing}.
They are consistent with $C_0=1,C_1=0$, i.e., the data suggest no modifications, and are perfectly consistent with the best-fit \lcdm+GR model.
It is interesting to notice that the preferred $\Omo$ is rather high with respect to the Planck value, while the preferred $\seighto$ is low. 
We checked that when going to higher orders in the crossing functions, i.e., including $C_2$ and $C_3$, the contours and do not suggest further modifications (i.e., is still consistent with $C_0=1, C_{i>0} = 0$). 

Fig.~\ref{fig:crossing} shows a random selection of 16 crossing functions and their effects on the reconstructed $\mu$ and on $(\Omo,\seighto)$. 
Their corresponding crossing parameters $C_i$ are shown as coloured points in Fig.~\ref{fig:mcmc_xing}.
This gives the reader some intuition on how distorting the growth affects the reconstructed expansion. 
Small distortions from the best-fit \lcdm+GR case can lead to significantly different $\hat\mu$.

\subsection{Gaussian Process regression}
\label{res:GP}
 
We used Gaussian Process regression \citep[GP,][]{2006gpml.book.....R} to reconstruct $\fseight(z)$ from the RSD measurements. 
GP have been increasingly used in cosmology \citep{2010PhRvD..82j3502H, 2010PhRvL.105x1302H, 2011PhRvD..84h3501H, 2012PhRvD..85l3530S, 2013PhRvD..87b3520S, 2018PhRvD..97l3501J, 2019MNRAS.485.2783L} and other fields of astronomy \citep[e.g.][]{2019arXiv190102877I}.
A Gaussian process is effectively a random sampling on a function space, generalizing random numbers.
{GP can be used to reconstruct a smooth function     $\vect{f}_*$ at the test points $\vect{x}_*$ given a discrete set of observations $(x_i,y_i)$ and a data covariance matrix $\tens{C}$.}
For a given kernel, the covariance between pairs of random variables \vect{u} and \vect{v} is thus given by
\tens{K}(f(\vect{u}),f(\vect{v})) = k(\vect{u},\vect{v}), where k(\vect{u},\vect{v}) is the covariance kernel. 
The joint-distribution of the training (observed) outputs \vect{y} and the test (reconstructed) output $\vect{f}_*$ is a Gaussian joint distribution given by
\begin{align}
  \begin{bmatrix}
	\vect{y} \\
    \vect{f}_*
  \end{bmatrix} & 
  	\sim \mathcal{N}\left(
  	\vect{0},
    \begin{bmatrix}
  		\tens{K}(X,X) + \tens{C} & \tens{K}(X,X_*)\\
  		\tens{K}(X_*,X)          & \tens{K}(X_*,X_*)
  	\end{bmatrix}
  \right)
\end{align}
where \tens{C} is the covariance of the data. 

\begin{figure*}
    \centering
    \includegraphics[width=\textwidth]{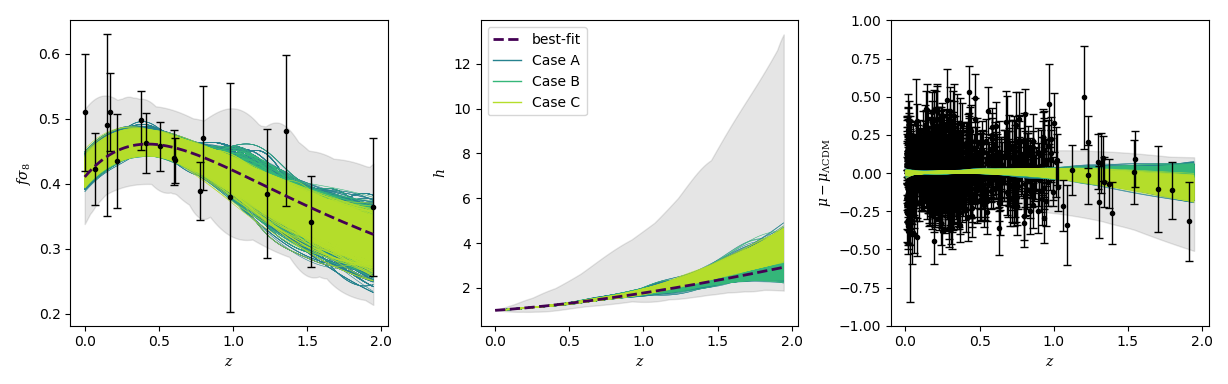}
    \caption{Left: \fseight, middle: reconstructed $h(z)$, Right: $\mu-\mu_\mathrm{\lcdm}$. 
    The grey shadow shows the envelope of all reconstructions, and the solid lines show the reconstructions (with $\chsqr < \chsqr_\mathrm{ref}$) for cases A, B, and C. 
    Dashed-lines: best-fit \lcdm\  (reference model).
    \label{fig:res_current}}
\end{figure*}

We use the squared exponential kernel defined as 
\begin{align}
k_{\sigma_f,\ell}(\vect{x},\vect{y}) & = \sigma_f^2\exp{\left(-\frac{\abs{\vect{x}-\vect{y}}^2}{2\ell^2}\right)},
\end{align}
where $(\sigma_f,\ell)$ are two hyperparameters controlling the amplitude and the correlation scale, and thus the deviation from the mean function.
For a given $(\sigma_f^2,\ell)$, we can thus generate a number of samples of $\fseight$ at any redshift $z$.

In practice, we start from the best-fit \lcdm\ as a mean function, use GP as a sampling of possible growth histories, and then apply the formalism from \S~\ref{sec:method}.

In order to prevent fitting the noise, we impose a hard prior on $\ell\in [0.2,1]$.

A notable difference with the work of \citet{2018arXiv180800377Y} is in the GP regression itself. 
While they obtain the mean and one-sigma contours, we follow each individual random sampling of the function space, therefore obtaining a set of plausible (and self-consistent) couples of expansion and growth histories.  
These two approaches are mathematically equivalent. However, in this case, since most of the constraining power comes from the SNIa, we do not train the GP on the \fseight\ data. This approach allow more flexibility in \fseight, which in turn allow more flexibility in the distance moduli.

\begin{figure*}
    \centering
    \includegraphics[width=\textwidth]{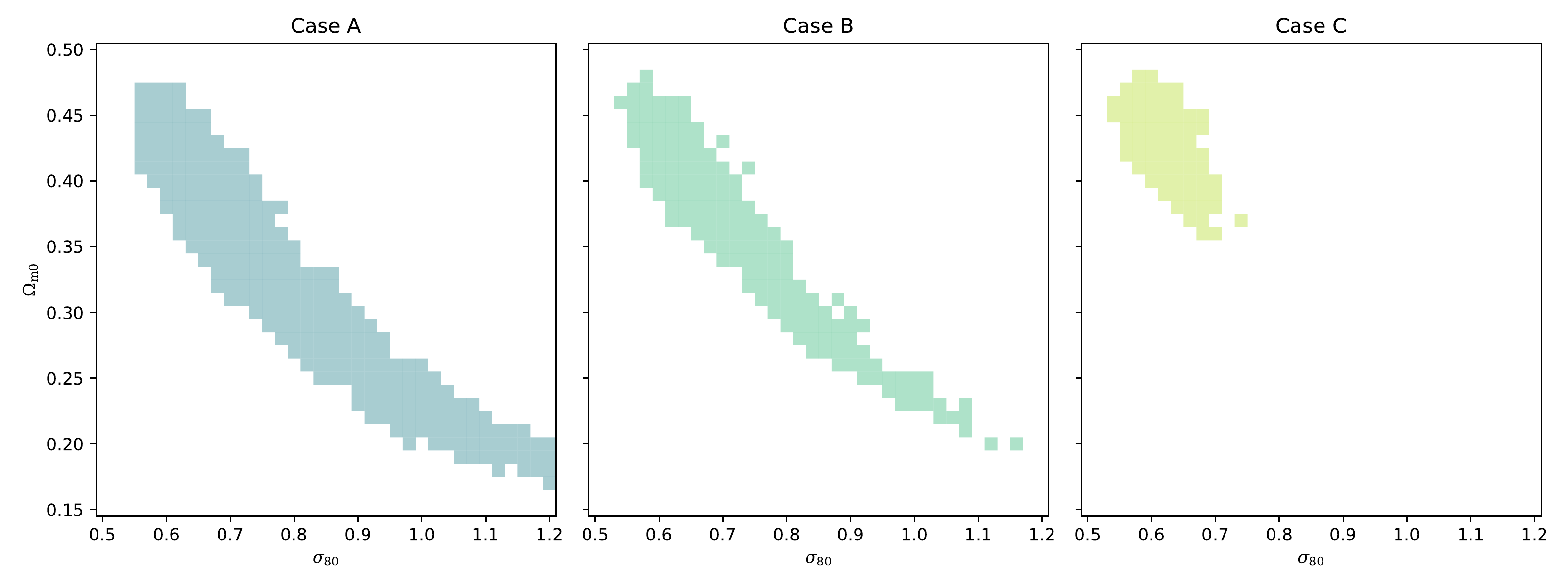}
    \caption{Model-independent constraints on $(\Omo,\seighto)$, that is, allowed contours in the $(\seighto,\Omo)$ plane for which we can find at least one $\fseightrec$ and its corresponding \hrec\ that fit the growth and SNIa better than \lcdm. 
    The left-hand, middle, and right-hand panels respectively show the contours for the reconstructions belonging to cases A, B, and C.
    }
    \label{fig:Oms8_current}
\end{figure*}

The left-hand panel of Fig.~\ref{fig:res_current} shows the reconstructed \fseightrec\ as well as the best fit in dashed lines.  
The shadowed area shows the envelope of the reconstructed \fseightrec.
A vast majority of these reconstructions do not fit the data ($\chsqr>\chsqr_\text{ref}$), therefore, we show in solid lines those reconstructions of \fseightrec\ that, together with some appropriate $(\Omo,\seighto)$, yields a $\chi^2_\text{tot} < \chsqr_\text{ref}$.
In order to study the effect of \zmax\ on the positive DE condition~\eqref{eq:pde}, we then separate the samples into three disjoint sets, according to the three values of $z_\text{max}$:
\begin{subequations}
\begin{align}
\mbox{Case A:} &\; \forall z \in [0,2] \quad   \Omde(z) >0,\\
\mbox {Case B:} &
    \begin{cases}
        \forall z \in [0,1] \quad \Omde(z)  >0 \\
        \exists z\in [1,2]  \quad \Omde(z)<0
    \end{cases}\\
    \mbox{Case C:}&
    \begin{cases}
        \forall z \in [0,0.7] \quad  \Omde(z)  >0 \\
        \exists z\in [0.7,1]  \quad \Omde(z)<0
    \end{cases} 
\end{align}
\end{subequations}
and show these in different colours. 
The middle and right-hand panels show the reconstructed $h(z)$ and $\mu-\mu_\text{ref}$, with the same convention.

Due to the oscillations in the reconstructed $\fseightrec$ from GP, the reconstructed shapes of $h$ and $\mu$ have more flexibility than \lcdm, yielding possible better fit to the data, therefore representing a non-exhaustive set of plausible expansion and growth histories. 
It is worth mentioning here that, since the process of reconstructing $h^2$ from $\fseightrec$ via eq.~\eqref{eq:delta2h2} involves two integrals, it is very sensitive to variations in the growth history and in the choice of $(\Omo,\seighto)$. 
In practice, most reconstructed $\fseightrec$ cannot yield any $\hrec$ that fits the SNIa data. 
This can be seen by the large grey envelope in the three panels of Fig.~\ref{fig:res_current} compared to the thinner band of allowed reconstructions. 
Therefore, it is important to explore the $(\sigma_f^2, \ell)$ parameter space and generate a large number of random realizations.

Fig.~\ref{fig:Oms8_current} shows the area of the $(\Omo,\seighto)$ parameter space in cases A, B, and C that, for at least one reconstructed $\fseightrec$, yields $\chsqr<\chsqr_\mathrm{ref}$, i.e., which is within the $1\sigma$ region of the best-fit \lcdm\ model.
In case A, the extent is maximal: for a large range of the parameters $(\Omo,\seighto)$, one may find reconstructions with $\chsqr<\chsqr_\text{ref}$ (and $\Omde>0$). 
As we start allowing $\Omde$ to become negative, the contours shrink. It is apparent in case C, where $\Omde$ crosses 0 between $z=0.7$ and 1, which is only possible for large values of \Omo\ (for a fixed $h(z)$, increasing \Omo\ allows \Omde\ to cross 0).  
It is worth noting that in cases B and C, \Omde\ crosses 0 at least once, but can become positive again. 

In Appendix~\ref{sec:mock}, we demonstrate the validity of the algorithm on a simulated data set, successfully recovering the input cosmology.

\begin{figure}
    \centering
    \includegraphics[width=\columnwidth]{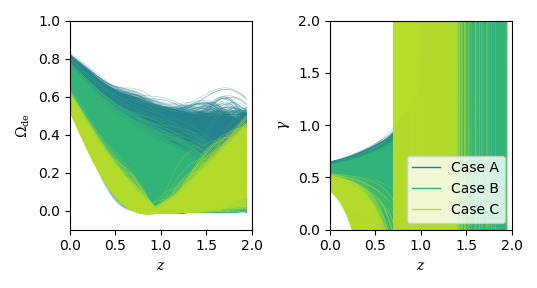}
    \caption{Reconstructions of $\Omde$ (left) and $\gamma$ (right). 
    The solid lines shows the reconstructions (with $\chsqr<\chi^2_\text{\lcdm}$) for cases A, B, and C.
    \label{fig:wgamma_current}}
\end{figure}

It is worth noting that, since we need to assume a value for $\Omo$ in order to reconstruct $h$, we can then obtain the (uniquely defined) equation of state $w(z)$ and growth rate $\gamma(z)$, as
\begin{align}
    w(z) & = \frac{\tfrac{2}{3}(1+z)\tfrac{h'}{h}-\Omm(z)}{1-\Omm(z)}-1,\qquad \mbox{and}\\
    \gamma (z) & = \frac{\ln f(z)}{\ln\Omm(z)}.
\end{align}  
We note that, since two consecutive integrations are needed to obtain $h$, then in theory only one is needed to obtain $h'$, making this method potentially less sensitive to numerical noise, provided that  the quality of the $f\sigma_8$  data improve substantially.
In addition, since the positive dark energy condition \eqref{eq:pde} ensures that $1-\Omm(z)\geq 0$ only up to \zmax, $w$ might diverge. 
For these reasons, instead of $w$, we reconstruct \Omde,
which does not involve derivatives of $h$. 
Fig.~\ref{fig:wgamma_current} shows our obtained reconstructions of $\Omde$ and $\gamma$.
All these reconstructions have $\chsqr<\chsqr_{\lcdm}$ and obey the positive dark energy condition for $z<\zmax$ for $\zmax=0.7,1,2$ as showed in different colours. 
The quality of current data does not constrain these functions beyond $z \gtrsim 0.1$. 
However, it is interesting that our reconstruction permits the DE energy density $\rho_\mathrm{de}$ to become negative (thus, $\Omm>1$) for $z\gtrsim 0.7$. This is not impossible and, in principle, may be realized by DE being a scalar field with a negative potential tracking dust-like matter for energy densities much exceeding the present critical one. On the other hand, it is clearly seen from Fig. 5 that the change of $\rho_\mathrm{de}$ from the positive present value to a negative one, if it occurs at all, can happen at sufficiently low redshifts $z<2$ only (the cases B and C).

It is interesting to compare these obtained model-independent constraints on $(\Omo,\seighto)$ (Fig.~\ref{fig:Oms8_current} with those obtained in \citet{2018PhRvD..98h3526S}, where the approach was different. 
In that paper, the authors reconstructed $h(z)$ from the Pantheon SNIa compilation via iterative smoothing, and then obtained $\fseightrec(z)$ by assuming a constant $\gamma = 0.55$ (or varying it as a free, but constant, parameter), and obtain $f=\Omm^\gamma(z)$.

\section{Summary \& Conclusion}
\label{sec:ccl}

Using the latest compilation of RSD measurements, we reconstruct the growth history using two model-independent approaches, namely, crossing statistics and Gaussian processes, only assuming a flat-FLRW Universe and general relativity. 
We then used the method introduced by \citet{1998JETPL..68..757S} to reconstruct the expansion history, and fit the corresponding distance moduli to the Pantheon SNIa compilation, and finally obtained model-independent constraints on $(\Omo,\seighto)$.
In addition, it is possible to reconstruct the dark energy equation of state $w(z)$ and the growth rate $\gamma(z)$. 
Applying the crossing statistics formalism, i.e.,  multiplying the best-fit \lcdm+GR growth by some hyperfunction, and obtained constraints on $\Omo,\seighto, C_i$, where the $C_i$ are hyperparameters of the crossing hyperfunctions, we find consistency with $C_0=1, C_i=0, i\geq 1$, i.e., the data do not call for any modification to the best-fit. 
However, the preferred values for $\Omo=0.381^{+0.049}_{-0.113}$ and $\seighto = 0.68^{+0.15}_{-0.12}$ are respectively higher and lower than the Planck best-fit, although consistent with them. 
Using Gaussian processes gives similar results. 
Both approaches suggest no {statistically significant} departure from \lcdm+GR. {On the other hand, it is interesting that they do not prohibit DE energy density from reaching zero at some recent redshift $0.7\lesssim z<2$ and becoming negative for larger $z$.}

Future surveys such as the Dark Energy Spectroscopic Instrument are expected to provide more accurate measurements of the growth, and thus, to further constrain the gravity model and its parameters.
This approach can be thought of as the reciprocal approach of \citet{2018PhRvD..98h3526S}, which uses direct reconstructions of $h$ from the supernovae data to fit the RSD data. 
Both approaches can be thought of as a mutual consistency test of the data and theory: in the former paper, the reconstructed expansion histories are tested against the growth data, while here, the reconstructed growth is compared to the SNIa data.

Our analyses point towards the consistency of the reconstructed \lcdm\ background evolution (via the mean function) with  the growth history inside GR.
While in GR, $\gamma = 0.55$ is a very good approximation as long as $\Omo$ is not too small (\citet{2016JCAP...12..037P}), it is not valid anymore beyond GR. 
On the contrary, the present approach, can be applied to non-GR models provided that $\Geff$ is known, or can even be used to reconstruct the effective Newton constant \Geff\ (L'Huillier et al. in prep), and therefore to constraining modified gravity.

\section*{Acknowledgements}
We thank the anonymous referee for suggesting to include negative DE energy densities in our study.
BL thanks Ryan~E.~Keeley for valuable comments about Gaussian Processes.
This work benefited from the Supercomputing Center/ Korea Institute of Science and Technology Information with supercomputing resources including technical support (KSC-2017-C2-0021) and the high performance computing clusters Polaris and Seondeok at the Korea Astronomy and Space Science Institute.
BL would like to acknowledge the support of the National Research Foundation of Korea (NRF-2019R1I1A1A01063740). 
A.A.S. was partly supported by the program KP19-270 ``Questions of the origin and evolution of the Universe'' of the Presidium of the Russian Academy of Sciences.




\bibliographystyle{mnras}
\bibliography{biblio} 


 
\appendix

\section{Validation on mock data}
\label{sec:mock}

\begin{figure*}
    \centering
    \includegraphics[width=\textwidth]{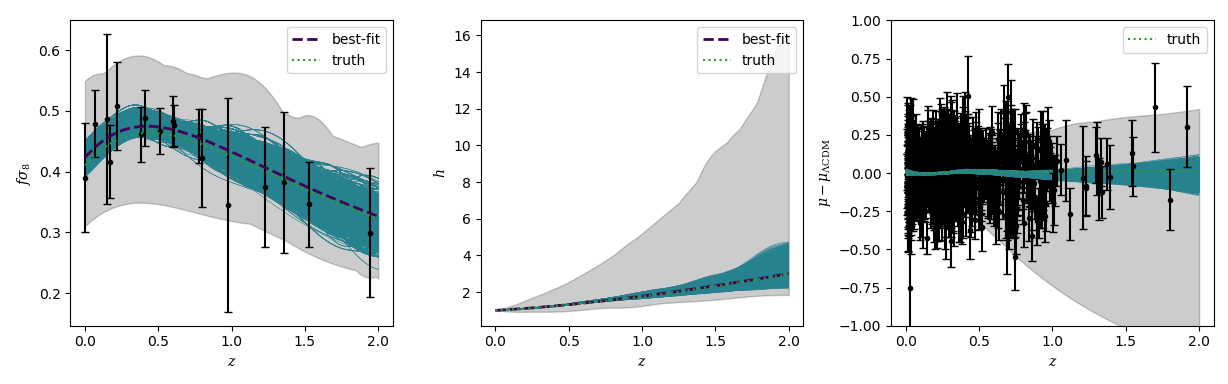}
    \caption{Reconstructions of $\fseight$ (left), $h^2$ (middle), and $\mu-\mu_\text{ref}$ (right) for the simulated data. 
    The shadowed area show the envelope of the reconstructions, and the solid lines are those reconstructions with yield a \chsqr\ better than $\chsqr_\mathrm{ref}$. 
    The best-fit \lcdm\ is shown in thick dashed lines, and the true cosmology in dotted lines.}
    \label{fig:res_sim}
\end{figure*}

\begin{figure}
    \centering
    \includegraphics[width=\columnwidth]{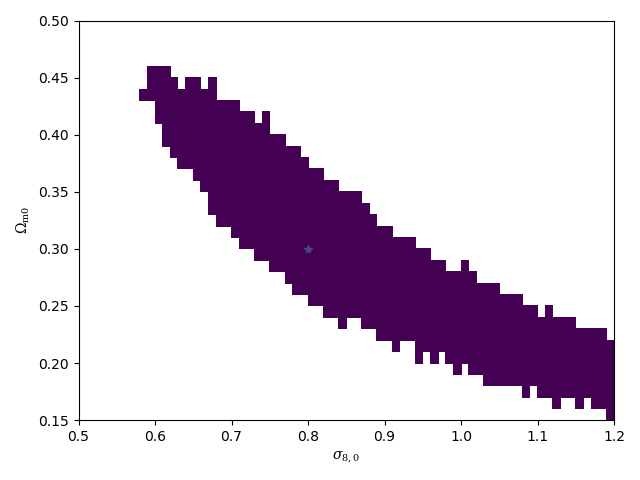}
    \caption{Allowed parameter space of $(\Omo,\seighto)$.
    All the points have a better \chsqr\ than the best-fit \lcdm\ model.
   }
    \label{fig:Oms8_sim}
\end{figure}

In order to validate the method, we applied it to a controlled simulated realization of the data, where the input cosmology is known. 
We generated mock data, following the redshift distribution and errors following the data in \citet{2018PhRvD..98h3526S}, assuming a known cosmology of $(\Omo,\seighto)=(0.3,0.8)$. 
We fit a flat-\lcdm\ universe to the total (RSD+SNIa) data, and use this best-fit \lcdm\ model as a mean function for the GP, and use its \chsqr\ as a reference. 
Hereafter, we use subscript $_\text{ref}$ to denote the best-fit model. 
We applied our pipeline, reconstruct \fseight\ and $h$, and calculate the $\chi^2$ to the data (mock growth and Pantheon-like SNIa).
Fig.~\ref{fig:res_sim} is the same as Fig.~\ref{fig:res_current} but with our simulated data, and without separating cases A, B, and C. 
In addition to the best-fit in dashed lines, the true cosmology is shown in dotted lines. 
Fig.~\ref{fig:Oms8_sim} shows the allowed contours for $(\Omo,\seighto)$, and the true cosmology, denoted by the green point, is inside the contours, showing the validity of the method.


\bsp	
\label{lastpage}
\end{document}